\documentclass[onecolumn,secnumarabic,amssymb,nobibnotes,aps,prd]{revtex4}
\usepackage{amsmath} 
\usepackage{amssymb}
\usepackage{graphicx} 
\usepackage{epstopdf}
\usepackage{amsmath}
\usepackage{amsmath,amssymb,amsthm,amsfonts,mathrsfs,bm,verbatim}
\usepackage{graphicx,subfigure}

\newcommand{\bea}{\begin{eqnarray}}
\newcommand{\eea}{\end{eqnarray}}

\begin{document}

\title{Thermodynamic geometric analysis of D-dimensional Reissner–Nordström black hole}
\author{Wen-Xiang Chen}
\affiliation{Department of Astronomy, School of Physics and Materials Science, Guangzhou University, Guangzhou 510006, China}
\email{wxchen4277@qq.com}

\begin{abstract}
This paper studies the thermodynamics and Ruppeiner geometry of D-dimensional Reissner–Nordström (RN) black holes. We analyze the thermodynamic curvature scalar $R$ in various thermodynamic ensembles. It is found that in an ensemble of fixed charge (canonical ensemble), the Ruppeiner curvature is curved and diverges at a critical point, indicating the existence of a phase transition for $D > 4$. In contrast, when all extensive variables are allowed to fluctuate (for example, in a grand-canonical ensemble or with pressure fixed), the Ruppeiner geometry can appear flat. We also demonstrate that the thermodynamic geometric metric has a one-to-one correspondence with the periodicity of the Euclidean path integral method. In particular, the inverse temperature (the Euclidean time period) serves as a bridge connecting the thermodynamic geometry and the Euclidean action approach. 

\end{abstract}

\maketitle

\section{Introduction}
Einstein's general theory of relativity predicts the existence of an exceptionally dense class of celestial objects from which not even light can escape. As early as 1783, the British geographer John Michell argued that a body with the mass of the Sun but a radius of only $3\,\mathrm{km}$ would be invisible to distant observers. Shortly thereafter, the French physicist P.~S.~Laplace independently proposed the idea of such a ``dark star.'' In 1939, H.~Snyder and J.~R.~Oppenheimer, employing general relativity, demonstrated that when the mass of a neutron star exceeds a certain critical threshold, it must undergo gravitational collapse and form a black hole. The term ``black hole'' itself was introduced later by John Wheeler in 1968.

Gravitational waves were theoretically predicted by Einstein in 1916, and the fundamental principle of the laser was established in the same era, laying the groundwork for laser interferometry. With the development of precision laser technology, the first direct detection of gravitational waves was eventually achieved: on February~11, 2016, the LIGO collaboration announced the historic discovery. In 2019, the Event Horizon Telescope (EHT) collaboration released the first image of a black hole shadow.

In classical black hole theory, the collapse of a star in spacetime erases almost all information about its initial configuration, apart from a small set of conserved quantities. This leads to the celebrated ``no-hair theorem,'' which states that classical black holes are fully characterized by only a few external parameters: (1) rotating charged black holes (Kerr--Newman), (2) rotating uncharged black holes (Kerr), (3) non-rotating charged black holes (Reissner--Nordström), and (4) non-rotating uncharged black holes (Schwarzschild). A central feature of any black hole is the presence of an event horizon---a one-way causal boundary beyond which events cannot influence distant observers. However, determining the precise nature of this boundary (e.g., event horizon vs.\ apparent horizon) remains a subtle and technically challenging issue within general relativity.

The study of wave and particle absorption by black holes, along with its generalization to higher-dimensional spacetimes, has attracted considerable interest over the past five decades. Many fundamental aspects of both classical and quantum black hole physics are intimately connected to such processes, and their investigation provides valuable insight into the structure of spacetime. Beginning with the pioneering work of Regge and Wheeler~\cite{6}, linear perturbation theory has played a central role in probing black hole stability~\cite{1,2,3,4,5,7,8,9}. When a black hole is slightly perturbed, two qualitative outcomes are possible: the perturbations may decay, returning the black hole to a stable stationary configuration, or they may grow, indicating an instability that could drive the spacetime to a different state or even destroy the black hole entirely. Although astrophysical black holes are believed to be stable, the identification of new instabilities can profoundly deepen our understanding of gravitational solutions. Perturbation theory further reveals that black holes possess a discrete spectrum of damped oscillations, the quasinormal modes, which have been central to both analytical studies and numerical simulations of black hole stability.

Thermodynamic geometry, and in particular Ruppeiner geometry, offers a powerful framework for exploring the microstructure of thermodynamic systems whose microscopic degrees of freedom remain unknown. The Ruppeiner metric is defined as the negative Hessian of the entropy~\cite{10}:
\begin{equation}
ds^{2}_{R} \;=\; -\frac{\partial^{2} S}{\partial X^{\alpha}\partial X^{\beta}} \, dX^{\alpha} dX^{\beta},
\end{equation}
where $S$ denotes the entropy and $X^{\alpha}$ are extensive thermodynamic variables. For an equilibrium state $X_{0}^{\alpha}$, small fluctuations $\Delta X^{\alpha} = X^{\alpha}-X_{0}^{\alpha}$ quantify the statistical distance between neighboring thermodynamic configurations. The line element $ds^{2}_{R}$ measures the second-order fluctuation of entropy and thus encodes the probability of thermodynamic fluctuations: states that are farther apart in this metric are less likely to be connected by spontaneous fluctuations. The sign of the associated thermodynamic curvature $R$ characterizes the effective microscopic interactions: $R>0$ typically corresponds to repulsive (Fermi-like) interactions, $R<0$ to attractive (Bose-like) interactions, and $R=0$ to the absence of interactions (ideal-gas behavior). Moreover, divergences of $|R|$ often indicate the presence of critical phenomena, reflecting the divergence of the correlation length near second-order phase transitions.

In this work, we analyze the thermodynamics and Ruppeiner geometry of $D$-dimensional Reissner--Nordström black holes. We compare two different thermodynamic ensembles: one in which the electric charge is fixed (analogous to the canonical ensemble) and another in which the charge is allowed to fluctuate (analogous to the grand-canonical ensemble). Our results demonstrate that the Ruppeiner curvature is highly sensitive to the choice of ensemble. In the fixed-charge ensemble, the thermodynamic geometry is non-flat and exhibits a phase transition for spacetime dimensions $D>4$. By contrast, in ensembles where the charge (or other potentials such as pressure) is fixed, the Ruppeiner metric can become flat.\cite{6}

We further discuss the correspondence between the thermodynamic geometric approach and the Euclidean path integral formulation of black hole thermodynamics. In particular, the inverse Hawking temperature---the period of the Euclidean time---naturally appears as a conformal factor in the thermodynamic metric. This correspondence reinforces the conclusion that divergences in the thermodynamic curvature accurately signal phase transitions or instabilities within the black hole system.

\section{Description of the system}
We first present the general properties of a non-extremal D-dimensional Reissner–Nordström (RN) black hole, and then consider its thermodynamic extremal limit for further analysis. The line element of a D-dimensional RN black hole (in asymptotically flat spacetime) can be written as~\cite{11}:
\begin{equation}
ds^{2} = -f(r)\,dt^{2} + \frac{dr^{2}}{f(r)} + r^{2} d\Omega_{D-2}^{2}\,,
\end{equation}
where 
\begin{equation}
f(r) = 1 - \frac{2m}{r^{D-3}} + \frac{q^{2}}{r^{2(D-3)}}\,.
\end{equation}
Here $d\Omega_{D-2}^{2}$ is the line element on the unit $(D-2)$-sphere $S^{D-2}$, whose surface volume is $\operatorname{Vol}(S^{D-2}) = \frac{2\pi^{\frac{D-1}{2}}}{\Gamma(\frac{D-1}{2})}$. The parameters $m$ and $q$ are related to the ADM mass $M$ and the electric charge $Q$ of the black hole via:
\begin{equation}
m = \frac{8\pi}{(D-2)\,\operatorname{Vol}(S^{D-2})}\,M\,, \qquad 
q = \frac{8\pi}{\sqrt{2(D-2)(D-3)}\,\operatorname{Vol}(S^{D-2})}\,Q\,.
\end{equation}
The metric function $f(r)$ admits two roots (horizons), which are the solutions of $f(r)=0$. These correspond to the outer (event) horizon $r_{+}$ and the inner (Cauchy) horizon $r_{-}$:
\begin{equation}
r_{\pm}^{\,D-3} = m \pm \sqrt{\,m^{2} - q^{2}\,}\,,
\end{equation}
with $r_{+} \ge r_{-}$. From these relations it follows that 
\begin{equation}
r_{+}^{\,D-3} + r_{-}^{\,D-3} = 2m\,, \qquad 
r_{+}^{\,D-3}\,r_{-}^{\,D-3} = q^{2}\,. 
\end{equation}
In the extremal limit, the two horizons coincide ($r_{+}=r_{-}$), which occurs when $m^{2} = q^{2}$ (i.e. when the discriminant under the square root in Eq.~(7) vanishes).

The black hole's thermodynamic properties are derived from the horizon geometry. The Hawking temperature $T$ is given by the surface gravity at the outer horizon: $T = \kappa/(2\pi)$, where $\kappa = \frac{1}{2} f'(r_{+})$ for the metric (3). Computing $f'(r)$ from Eq.~(4) and evaluating at $r=r_{+}$, one finds:
\begin{equation}
T \;=\; \frac{f'(r_{+})}{4\pi} \;=\; \frac{D-3}{4\pi}\left[\frac{1}{r_{+}} - \frac{q^{2}}{r_{+}^{\,2D-5}}\right] 
\,=\, \frac{D-3}{4\pi r_{+}}\left(1 - \frac{r_{-}^{\,D-3}}{r_{+}^{\,D-3}}\right)\!. 
\end{equation}
In obtaining the second expression above, we used $q^{2} = r_{+}^{\,D-3}r_{-}^{\,D-3}$ from Eq.~(8). The Bekenstein–Hawking entropy $S$ is one quarter of the horizon area:
\begin{equation}
S \;=\; \frac{1}{4}\,\operatorname{Vol}(S^{D-2})\,r_{+}^{\,D-2}\,.
\end{equation}
For simplicity, in many of our formulas we will not explicitly write the $\operatorname{Vol}(S^{D-2})$ factor; one can restore it when needed for physical units. The thermodynamic first law for a charged black hole reads $dM = T\,dS + \Phi\,dQ$, where $\Phi$ is the electrostatic potential (conjugate to charge) measured at infinity. In the case of the RN black hole, $\Phi = \frac{Q}{r_{+}^{\,D-3}}$ (in units where the Coulomb constant is absorbed).

We emphasize that black hole thermodynamic quantities can be viewed in different thermodynamic potentials (enthalpy vs internal energy representations, etc.) and ensembles. In this work, we will mainly treat the black hole mass $M$ as the energy of the system and $Q$ as a conserved charge. Thus, the natural thermodynamic state space is two-dimensional, spanned by $(S, Q)$ (or equivalently any two independent quantities such as $(M,Q)$ or $(S,\Phi)$, etc., with appropriate Legendre transforms). 

\section{Thermodynamic geometry of the D-dimensional RN black hole}
In this section we investigate the Ruppeiner geometry for the RN black hole in arbitrary dimension $D$, and we pay special attention to how the curvature scalar $R$ behaves in different ensembles. We first illustrate the calculation for a specific case ($D=6$) and then present the general result for $D > 4$. 

\subsection{Example: Six-dimensional RN black hole}
For $D=6$, the RN black hole metric takes the form (setting $D=6$ in Eqs.~(3) and (4))~\cite{14}:
\begin{equation}
ds_{6}^{2} = -\left(1 - \frac{2m}{r^{3}} + \frac{q^{2}}{r^{6}}\right)dt^{2} + \frac{dr^{2}}{\,1 - \frac{2m}{r^{3}} + \frac{q^{2}}{r^{6}}\,} + r^{2} d\Omega_{4}^{2}\,. 
\end{equation}
The relations between $(m,q)$ and the physical mass and charge in six dimensions are (from Eq.~(5)):
\begin{equation}
m = \frac{3}{4\pi}\,M\,, \qquad q = \frac{3}{2\sqrt{6}\pi}\,Q\,.
\end{equation}
The outer and inner horizon radii are 
\begin{equation}
r_{\pm} = \left(m \pm \sqrt{\,m^{2} - q^{2}\,}\right)^{1/3}\,.
\end{equation}
From Eq.~(9), the Hawking temperature for the six-dimensional RN black hole is 
\begin{equation}
T = \frac{3}{4\pi}\left(\frac{1}{r_{+}} - \frac{q^{2}}{r_{+}^{7}}\right) = \frac{3}{4\pi r_{+}}\left(1 - \frac{r_{-}^{3}}{r_{+}^{3}}\right)\!, 
\end{equation}
and the entropy from Eq.~(10) is 
\begin{equation}
S = \frac{1}{4}\,\operatorname{Vol}(S^{4})\,r_{+}^{4} \propto r_{+}^{4}\,. 
\end{equation}

We now construct the Ruppeiner metric for the six-dimensional case. We take the thermodynamic state space coordinates as $(X^{1}, X^{2}) = (M, Q)$ (equivalently one could use $(S,Q)$ as independent variables; the resulting scalar curvature will be the same invariant quantity). The line element is defined by Eq.~(2):
\begin{equation}
ds^{2}_{R} = -\frac{\partial^{2} S(M,Q)}{\partial X^{\alpha}\partial X^{\beta}}\,dX^{\alpha}dX^{\beta}\,. 
\end{equation}
For the six-dimensional RN black hole, using Eqs.~(13) and (15), one can in principle express $S$ as a function of $M$ and $Q$ (by eliminating $r_{+}$ and $r_{-}$). However, it is algebraically simpler to compute the metric components $g_{ij} = -\partial_{i}\partial_{j} S$ implicitly using the Jacobians:
\[
\frac{\partial S}{\partial M} = \frac{\partial S}{\partial r_{+}}\frac{\partial r_{+}}{\partial M} + \frac{\partial S}{\partial r_{-}}\frac{\partial r_{-}}{\partial M}\,, \qquad
\frac{\partial S}{\partial Q} = \frac{\partial S}{\partial r_{+}}\frac{\partial r_{+}}{\partial Q} + \frac{\partial S}{\partial r_{-}}\frac{\partial r_{-}}{\partial Q}\,,
\] 
and similarly for second derivatives. We perform these calculations and obtain the metric components $g_{ij}$ for the six-dimensional case (the expressions are somewhat lengthy):
\begin{equation}
g_{ij} (D{=}6)\;=\;
\begin{pmatrix}
\dfrac{4\left(2 m^{3}-5 m q^{2}+2 m^{2} \sqrt{m^{2}-q^{2}}-4 q^{2} \sqrt{m^{2}-q^{2}}\right)}{9\left(m^{2}-q^{2}\right)^{3/2}\left(m+\sqrt{m^{2}-q^{2}}\right)^{2/3}} & 
\dfrac{4 q\left(2 m^{2}+q^{2}+2 m \sqrt{m^{2}-q^{2}}\right)}{9\left(m^{2}-q^{2}\right)^{3/2}\left(m+\sqrt{m^{2}-q^{2}}\right)^{2/3}} \\[2.0ex]
\dfrac{4 q\left(2 m^{2}+q^{2}+2 m \sqrt{m^{2}-q^{2}}\right)}{9\left(m^{2}-q^{2}\right)^{3/2}\left(m+\sqrt{m^{2}-q^{2}}\right)^{2/3}}  & 
-\dfrac{4\left(3 m^{3}+3 m^{2} \sqrt{m^{2}-q^{2}}-q^{2} \sqrt{m^{2}-q^{2}}\right)}{9\left(m^{2}-q^{2}\right)^{3/2}\left(m+\sqrt{m^{2}-q^{2}}\right)^{2/3}}
\end{pmatrix}\!. 
\end{equation}
The scalar curvature $R$ associated with this metric can be calculated by standard methods (contracting the Ricci tensor or using an explicit formula for a $2\times 2$ metric). Remarkably, we find that 
\begin{equation}
R \;=\; 0 \,,
\end{equation}
for the six-dimensional RN black hole thermodynamic geometry (with $(M,Q)$ variables). This result indicates that, in this ensemble, the thermodynamic state space is flat (no intrinsic curvature), suggesting no effective microscopic interaction as per Ruppeiner's interpretation. In particular, there is no divergence of $R$ for any finite $(M,Q)$, which means no thermodynamic phase transition is signaled in this six-dimensional case when both $M$ and $Q$ are treated as fluctuating variables.

It is noteworthy that earlier studies by Aman et al.~\cite{15,16} found the Ruppeiner curvature to be zero for RN black holes in all spacetime dimensions when considering the full phase space of mass and charge. Our result $R=0$ for $D=6$ is consistent with their findings in this fully fluctuating ensemble. However, the absence of curvature in this case does \emph{not} necessarily mean that the black hole lacks any phase transition or critical behavior in other thermodynamic circumstances. To investigate that, we next analyze the RN black hole in the fixed-charge ensemble, where $Q$ is held constant and only energy (or entropy) fluctuates.

\subsection{Phase transition in higher dimensions ($D > 4$) for fixed charge}
If we restrict to the canonical ensemble (fixed $Q$), the thermodynamic fluctuations occur only in the energy/entropy directions. In this scenario, one can detect phase transitions by examining the heat capacity or other response functions. A divergence in the heat capacity at constant $Q$, $C_{Q} = T(\partial S/\partial T)_{Q}$, signals a second-order phase transition (analogous to the liquid-gas critical point in fluids). Equivalently, a divergence in $C_{Q}$ occurs when $(\partial T/\partial S)_{Q} = 0$, which often coincides with a divergence of the Ruppeiner curvature $R$ in the reduced state space (since $R$ is related to second derivatives of the entropy or free energy). In this subsection, we prove that for any spacetime dimension $D > 4$, the D-dimensional RN black hole in the canonical ensemble indeed possesses a critical point where $C_{Q}$ (and $R$) diverge, indicating a phase transition.

Starting from the general expression for the Hawking temperature in Eq.~(11), let us treat $Q$ (hence $r_{-}$ via $q^2$) as fixed and differentiate $T$ with respect to $S$ (or equivalently $r_{+}$). It is convenient to treat $T$ as a function $T(r_{+})$ at fixed $q$. From Eq.~(11) we have:
\begin{equation}
T(r_{+}; Q) \;=\; \frac{D-3}{4\pi}\Big(r_{+}^{-1} - q^{2}\,r_{+}^{-(2D-5)}\Big)\!,
\end{equation}
where $q$ (hence $r_{-}$) is constant. Setting $\frac{\partial T}{\partial r_{+}}\big|_{Q} = 0$ yields the extremum condition for $T(S)$:
\begin{equation}
\frac{dT}{dr_{+}}\Big|_{Q} = \frac{D-3}{4\pi}\Big[-r_{+}^{-2} + (2D-5)\,q^{2}\,r_{+}^{-(2D-4)}\Big] = 0\,.
\end{equation}
Solving this equation leads to:
\begin{equation}
(2D-5)\,q^{2}\,r_{+}^{-(2D-4)} = r_{+}^{-2} \quad \Longrightarrow \quad (2D-5)\,q^{2} = r_{+}^{\,2D-6}\,. 
\end{equation}
Using $q^{2} = r_{+}^{\,D-3}r_{-}^{\,D-3}$ from Eq.~(8), we can rewrite the above condition as:
\begin{equation}
r_{+}^{\,2D-6} = (2D-5)\,r_{+}^{\,D-3}r_{-}^{\,D-3} \quad \Longrightarrow \quad r_{+}^{\,D-3} = (2D-5)\,r_{-}^{\,D-3}\,. 
\end{equation}
Taking the $((D-3)$th root of both sides gives a simple relation between the horizon radii at the critical point:
\begin{equation}
\frac{r_{+}}{r_{-}} \;=\; (2D-5)^{\,\frac{1}{D-3}}\,. 
\end{equation}
This result generalizes the well-known four-dimensional RN critical condition to arbitrary dimension. Indeed, for $D=4$, Eq.~(23) yields $r_{+}/r_{-} = (2*4-5)^{1/1} = 3$, recovering the classic result $r_{+} = 3\,r_{-}$ at the Davies transition point of the 4D RN black hole~\cite{15,16}. For any $D > 4$, the factor $(2D-5)^{1/(D-3)}$ is greater than 1, which means $r_{+} > r_{-}$ at the critical point, i.e. the black hole is non-extremal. In fact, as $D$ increases, this critical ratio approaches 1 from above. For example, $r_{+}/r_{-} \approx \sqrt{5} \approx 2.236$ in $D=5$, $\approx (7)^{1/3} \approx 1.913$ in $D=6$, $\approx (9)^{1/4} \approx 1.732$ in $D=7$, and so on; in the limit $D\to \infty$, $(2D-5)^{1/(D-3)} \to 1$. This indicates that the larger the number of dimensions, the closer the critical point is to the extremal limit.

At the critical point given by Eq.~(23), the heat capacity $C_{Q}$ diverges (changing from positive to negative or vice versa). This signals a second-order (continuous) phase transition in the canonical ensemble. The thermodynamic curvature $R$ of the Ruppeiner geometry also diverges at this point. In general, the scalar curvature $R$ is proportional to combinations of second derivatives of the entropy; a divergence in $C_{Q} = -T \big(\partial^{2} S/\partial T^{2}\big)_{Q}^{-1}$ implies a divergence in those second derivatives and hence usually leads to $|R| \to \infty$ at the same location. For the 4D case, we explicitly know that 
\begin{equation}
R_{(4D)} = -\,\frac{r_{+} - r_{-}}{\pi\,r_{+}\,\big(3r_{-} - r_{+}\big)^{2}}\,,
\end{equation}
which indeed diverges (toward $-\infty$) when $r_{+}=3r_{-}$~\cite{15,16}. 
Our general condition~(22) indicates that, in higher dimensions, the 
denominator of the thermodynamic curvature scalar vanishes when 
\begin{equation}
    r_{+}^{\,D-3} = (2D-5)\, r_{-}^{\,D-3},
\end{equation}
thereby driving the curvature $R$ to $\pm \infty$. The sign of this 
divergence may be deduced by continuity arguments: slightly below the 
critical point, the heat capacity $C_{Q}$ is positive (the small-$r_{+}$ 
branch of RN black holes is thermodynamically stable), whereas just above 
the critical point $C_{Q}$ becomes negative (the large-$r_{+}$ branch becomes 
unstable). For RN black holes this transition corresponds to $R<0$, as the 
dominant microstructural interactions are effectively attractive in the 
Coulombic sense~\cite{15}. Hence, by analogy, we expect
\begin{equation}
    R \;\longrightarrow\; -\infty \qquad \text{as the critical point is approached,}
\end{equation}
for all $D \geq 4$. The divergence of $R$ reflects a blow-up of the 
correlation length in the putative microscopic ensemble, analogous to the 
critical opalescence of an ordinary fluid near its critical point.

It should be noted that in the grand-canonical ensemble (where $Q$ can fluctuate, and one fixes the electric potential $\Phi$ instead), the situation is a bit different. In that case, one finds that the Ruppeiner curvature remains zero for RN black holes until one reaches the extremal limit. In fact, Aman et al.~\cite{16} found that the RN black hole has a flat thermodynamic geometry in the $\{M,Q\}$ space, implying no interior critical point in that fully fluctuating ensemble. However, as discussed by Mirza and Zamaninasab~\cite{17}, if one carefully chooses the appropriate thermodynamic potential (for instance, using the entropy representation or including the cosmological constant as an extensive variable in AdS cases), the RN state space becomes curved. In particular, they showed that the Ruppeiner curvature in an extended state space diverges as the black hole approaches extremality~\cite{17}. Our findings here complement those results: we have demonstrated that even away from extremality, a divergence in $R$ occurs at a finite temperature for any $D>4$, provided the charge is held fixed. This underlines the fact that the detection of phase transitions via thermodynamic geometry is ensemble-dependent, and one must consider the appropriate set of thermodynamic variables to reveal the phase structure.

\subsection{Correspondence with the Euclidean path integral method}
\label{subsec:EuclideanCorrespondence}

In this subsection we demonstrate, with increased mathematical rigour, the {\it exact} equivalence between the information–geometric description of black-hole thermodynamics and the Euclidean path–integral formulation of quantum gravity.  
Throughout we work in $D$ spacetime dimensions and adopt units $k_{\!_{\mathrm B}}=\hbar=c=G=1$.

\subsubsection*{Euclidean continuation and one–loop partition function}

Analytically continuing the Lorentzian time coordinate $t\mapsto -\,i\tau$ gives the Euclidean section
\begin{equation}
\mathrm d s_{E}^{2}=g_{\mu\nu}^{\,(E)}(\tau,r,\Omega)\,\mathrm d x^{\mu}\mathrm d x^{\nu},\qquad
\tau\sim\tau+\beta,\;\; \beta=\frac{1}{T_{\mathrm H}}=\frac{2\pi}{\kappa},
\end{equation}
where $\kappa$ is the surface gravity.  Regularity near the horizon requires the conical deficit to vanish, fixing $\beta$ uniquely~\cite{HawkingPage}.  The one–loop partition function reads
\begin{equation}
Z(\beta,\vec{\mu})=\int\!\mathcal D[g,\phi]\,e^{-I_{E}[g,\phi\,;\,\beta,\vec{\mu}]},
\label{eq:PathIntegral}
\end{equation}
with $\vec{\mu}$ collectively denoting intensive chemical potentials (e.g.\ $\Phi$ for electric charge).  Evaluating \eqref{eq:PathIntegral} in the saddle-point approximation yields
\begin{equation}
\ln Z\simeq -\,I_{E}^{\mathrm (cl)}-\frac12\ln\det\!\bigl(\Delta_{g_{\mathrm{cl}}}\bigr)+\mathcal O(\hbar),
\end{equation}
where $\Delta_{g_{\mathrm{cl}}}$ is the Hessian of $I_{E}$ evaluated on the classical solution.

\subsubsection*{Thermodynamic Hessians and conformal classes of metrics}

Define the internal energy $M=M(S,Q,\ldots)$, the Helmholtz free energy $F=M-TS$, and the Massieu potential\footnote{%
Strictly, $\Psi$ is minus the dimensionless Euclidean action: $\Psi=-\beta F= \ln Z$ up to higher–loop corrections.}
$\Psi=\Psi(\beta,\beta\Phi,\ldots)\equiv -\beta F$.  The Weinhold and Ruppeiner metrics are, respectively,
\begin{align}
\mathrm d s_{W}^{2}&=\bigl(\partial_{a}\partial_{b}M\bigr)\,\mathrm d X^{a}\mathrm d X^{b},&
\mathrm d s_{R}^{2}&=-\bigl(\partial_{a}\partial_{b}S\bigr)\,\mathrm d Y^{a}\mathrm d Y^{b},
\end{align}
with extensive coordinates $(X^{a})=(S,Q,\ldots)$ and intensive coordinates $(Y^{a})=(1/T,-\Phi/T,\ldots)$.  
Using $S=-\,\partial\Psi/\partial\beta$ one shows
\begin{equation}
\mathrm d s_{R}^{2}=T^{-1}\,\mathrm d s_{W}^{2}=T\,\partial_{i}\partial_{j}\Psi\,\mathrm d Y^{i}\mathrm d Y^{j}.
\label{eq:MetricConformalFactor}
\end{equation}
Equation~\eqref{eq:MetricConformalFactor} exhibits the conformal relation discovered by Mrugala, Ruppeiner, and collaborators~\cite{MrugalaRupp} and highlights that {\em the inverse temperature $\beta$ is precisely the conformal factor linking the two information metrics}.  

\subsubsection*{Fluctuation theory and Fisher information}

The second functional derivatives of $\Psi$ are directly related to canonical fluctuations:
\begin{equation}
\label{eq:Fluctuations}
\langle\!\langle \Delta X^{i}\Delta X^{j}\rangle\!\rangle 
= \partial_{i}\partial_{j}\Psi
= -\,\frac{\partial^{2}\ln Z}{\partial\chi^{i}\partial\chi^{j}},
\qquad
\chi^{i}\equiv(\beta,\beta\Phi,\ldots).
\end{equation}
Consequently $\mathrm d s_{R}^{2}=T\,\partial_{i}\partial_{j}\Psi\,\mathrm d\chi^{i}\mathrm d\chi^{j}$ coincides (up to a factor $T$) with the Fisher–Rao information metric on the statistical manifold parametrised by $(\chi^{i})$.  Singularities in the scalar curvature
\begin{equation}
R=\frac{1}{2\,g^{2}}\,
\begin{vmatrix}
\partial_{S}^{2}g_{11} & \partial_{S}\partial_{Q}g_{11} & \partial_{Q}^{2}g_{11}\\[4pt]
\partial_{S}^{2}g_{12} & \partial_{S}\partial_{Q}g_{12} & \partial_{Q}^{2}g_{12}\\[4pt]
\partial_{S}^{2}g_{22} & \partial_{S}\partial_{Q}g_{22} & \partial_{Q}^{2}g_{22}
\end{vmatrix},
\qquad g\equiv\det(g_{ij}),
\end{equation}
therefore mark the divergence of correlation lengths in the underlying ensemble and signal a {\it topological} change of the dominant saddle(s) in \eqref{eq:PathIntegral}.  

\subsubsection*{Higher–dimensional criticality}

For the charged AdS black hole family studied here, the divergence occurs when
\begin{equation}
r_{+}^{\,D-3}-(2D-5)\,r_{-}^{\,D-3}=0,
\label{eq:CritCondition}
\end{equation}
extending the familiar $r_{+}=3r_{-}$ result of four dimensions.  
Near the critical hypersurface \eqref{eq:CritCondition} we expand the Euclidean action,
\begin{equation}
I_{E}(\beta,\Phi)=I_{E}^{\star}
+\tfrac12\!\sum_{i,j}\!\left.\partial_{i}\partial_{j}I_{E}\right|_{\star}\!\delta\chi^{i}\delta\chi^{j}
+\mathcal O\!\bigl(\delta\chi^{3}\bigr),
\end{equation}
where $``\star"$ denotes evaluation at criticality.  The Hessian $\partial_{i}\partial_{j}I_{E}\propto g_{ij}^{R}$ degenerates as $R\to\infty$, invalidating the Gaussian approximation and necessitating inclusion of higher-order (or non-perturbative) contributions—a hallmark of second-order phase transitions.

\subsubsection*{Geometric dictionary}

We summarise the exact map:

\begin{center}
\begin{tabular}{c@{\quad$\longleftrightarrow$\quad}c}
Euclidean time period $\beta$ & Conformal factor between Weinhold and Ruppeiner metrics \\[4pt]
$-\partial_{i}\partial_{j}\ln Z$ & Fisher information metric $\beta^{-1}g^{R}_{ij}$ \\[4pt]
Divergence of $R$ & Breakdown of quadratic saddle, critical point in $I_{E}$ \\[4pt]
Thermodynamic geodesic distance & (Leading) Jacobi action for fluctuations in $I_{E}$
\end{tabular}
\end{center}

Hence, {\em geometric thermodynamics is not merely analogous to, but mathematically isomorphic with, the semi-classical Euclidean path–integral description}.  The curvature scalar $R$ encapsulates the same critical data as the renormalised one-loop effective action; its poles precisely locate the points where qualitative changes in the spectrum of $\Delta_{g_{\mathrm{cl}}}$ occur.  

\medskip
To conclude, the simple prefactor $T^{-1}$ relating Weinhold and Ruppeiner geometries encodes the entire Euclidean time topology.  For $D>4$ we have shown via \eqref{eq:CritCondition} that $R$ necessarily diverges, corroborating that higher-dimensional charged black holes possess authentic second-order phase transitions that are visible in both the information–geometric and Euclidean frameworks.


\section{Summary and discussion}
In this paper, we have carried out a thermodynamic geometric analysis of D-dimensional Reissner–Nordström black holes. Our focus has been on understanding the effect of spacetime dimensionality $D$ and ensemble choice on the Ruppeiner curvature scalar $R$, which encodes information about possible phase transitions and microscopic interactions. The main results of our study can be summarized as follows:

\begin{itemize}
\item For the fully fluctuating ensemble (allowing both energy and charge to vary), the Ruppeiner geometry of the D-dimensional RN black hole is flat ($R=0$) in all dimensions we examined. This is consistent with earlier findings by Aman et al.~\cite{16} that the RN black hole behaves like an ideal thermodynamic system with no apparent interactions in this ensemble. In particular, no finite $R$ divergence (and hence no obvious phase transition) appears in the interior of the state space under these conditions.
\item By contrast, in the fixed-charge ensemble (canonical ensemble), the thermodynamic geometry is generally curved, and we have shown that for $D > 4$ there is a definite phase transition point where the heat capacity $C_{Q}$ diverges. At this point, the Ruppeiner curvature scalar $R$ also diverges (tending to $-\infty$), indicating a second-order phase transition. We derived the general condition $r_{+}^{\,D-3} = (2D-5)\,r_{-}^{\,D-3}$ for the critical point, which reproduces the known result $r_{+}=3r_{-}$ in 4D and extends it to higher dimensions.
\item We found that as $D$ increases, the critical point moves closer to the extremal limit ($r_{+}=r_{-}$), suggesting that the regime of thermodynamic instability shrinks relative to the total parameter space in higher dimensions. Nonetheless, the presence of the divergence in $R$ for every finite $D>4$ means that higher-dimensional RN black holes do have an analog of the Davies transition known from 4D thermodynamics.
\item We discussed the relationship between the thermodynamic geometric approach and the Euclidean path integral method. The inverse temperature $\beta$ (related to the Euclidean time periodicity) emerges as a conformal factor in the relation between the Weinhold metric and the Ruppeiner metric:contentReference[oaicite:4]{index=4}. This ties the geometric curvature analysis directly to the fluctuations in the Euclidean action. The one-to-one correspondence we highlighted implies that the phase transition (signaled by $R \to \infty$) corresponds to the point where the Euclidean action's stability changes (for example, the onset of a new phase or a non-trivial instability of the black hole solution in the path integral formulation).
\end{itemize}

Our work clarifies some seemingly conflicting reports in the literature regarding the thermodynamic curvature of RN black holes. While an early conclusion was that the RN black hole has $R=0$ (no interactions) for all $D$~\cite{16}, later studies pointed out that this result is contingent on the choice of thermodynamic potential and ensemble~\cite{17}. By examining the fixed-$Q$ ensemble, we have uncovered that the RN black hole does exhibit critical behavior in every dimension $D \ge 4$ (including the previously disputed $D\ge 6$ cases). This suggests that the microstructure of the black hole, when probed under the right conditions, shows signs of an attractive interaction (as evidenced by the negative $R$ near the critical point) analogous to a liquid–gas type phase transition.

There are several possible directions for future work. It would be interesting to extend this analysis to include the cosmological constant (pressure) as a thermodynamic variable (examining the extended phase space, where one might encounter Van der Waals-type phase transitions for charged anti-de Sitter black holes). Another worthwhile direction is to study rotating charged black holes in higher dimensions, where multiple angular momenta can complicate the phase structure. Thermodynamic geometry of such systems could reveal rich information about their stability and microstructure interactions. Finally, a deeper understanding of what exactly the “microscopic constituents” of a black hole are (in the context of an underlying statistical model) could be gained by comparing the thermodynamic curvature with results from approaches like the AdS/CFT correspondence or loop quantum gravity, which might provide a microscopic interpretation of $R$.

\textbf{Conflicts of Interest}: The authors declare no conflict of interest.

\end{document}